\documentclass[prl,aps,twocolumn,superscriptaddress]{revtex4-1}
\usepackage{amssymb,amsfonts,amsmath,graphicx,color,times,mathtools}
 \pdfoutput=1

\def\tr{\mbox{tr}}
\def\bra#1{\langle{#1}|}
\def\ket#1{|{#1}\rangle}

{\catcode`\|=\active 
  \gdef\Braket#1{\begingroup
\mathcode`\|32768\let|\BraVert\left<{#1}\right>\endgroup}}
\def\BraVert{\egroup\,\mid\,\bgroup}

\newcommand{\Z}{\mathbb{Z}}

\DeclareMathOperator{\circula}{circ}
\DeclareMathOperator{\diag}{diag}
\DeclareMathOperator{\sgn}{sgn}

\newcommand{\T}{T}
\newcommand{\Taverage}{\overline{\T}}

\newcommand{\e}{\mathrm{e}}

\definecolor{christian}{rgb}{0.2,0.2,1.0}

\definecolor{jens}{rgb}{0,0.5,0.9}

\definecolor{antonello}{rgb}{0.1,0.5,0.1}

\definecolor{alessandro}{rgb}{0.5,0.1,0.1}

\definecolor{john}{rgb}{1,0.1,0.1}

\definecolor{stephen}{rgb}{0.8,0.1,0.5}

\allowdisplaybreaks

\begin{document}

\title{Total correlations of the diagonal ensemble herald the many-body localization transition}

\author{J.\ Goold}
\email{jgoold@ictp.it}
\affiliation{The Abdus Salam International Centre for Theoretical Physics (ICTP), Trieste, Italy}

\author{C.\ Gogolin}
\email{publications@cgogolin.de}
\affiliation{ICFO-The Institute of Photonic Sciences, Mediterranean Technology Park, 08860 Castelldefels (Barcelona), Spain}
\affiliation{Max-Planck-Institut f{\"u}r Quantenoptik, Hans-Kopfermann-Stra{\ss}e 1, 85748 Garching, Germany}

\author{S.\ R.\ Clark}
\email{s.clark1@physics.ox.ac.uk}
\affiliation{Department of Physics, Oxford University, Clarendon Laboratory, Parks Road, Oxford, UK}

\author{J.\ Eisert}
\email{jense@physik.fu-berlin.de}
\affiliation{Dahlem Center for Complex Quantum Systems, Freie Universit\"at Berlin, 14195 Berlin, Germany}

\author{A.\ Scardicchio}
\email{ascardic@ictp.it}
\affiliation{The Abdus Salam International Centre for Theoretical Physics (ICTP), Trieste, Italy}
\affiliation{INFN, Sezione di Trieste, I-34151, Trieste, Italy}

\author{A.\ Silva}
\email{asilva@sissa.it}
\affiliation{SISSA-International School for Advanced Studies, via Bonomea, 265, 34136 Trieste, Italy}
\affiliation{The Abdus Salam International Centre for Theoretical Physics (ICTP), Trieste, Italy}

\date{\today}

\begin{abstract}
The intriguing phenomenon of many-body localization (MBL) has attracted significant interest recently, but a complete characterization is still lacking.
In this work we introduce the total correlations, a concept from quantum information theory capturing multi-partite correlations, to the study of this phenomenon.
We demonstrate that the total correlations of the diagonal ensemble provides a meaningful diagnostic tool to pin-down, probe, and better understand the MBL transition and ergodicity breaking in quantum systems.
In particular, we show that the total correlations has sub-linear dependence on the system size in delocalized, ergodic phases, whereas we find that it scales extensively in the localized phase developing a pronounced peak at the transition.
We exemplify the power of our approach by means of an exact diagonalization study of a Heisenberg spin chain in a disordered field. By a finite size scaling analysis of the peak position and crossover point from log to linear scaling we collect evidence that ergodicity is broken before the MBL transition in this model.
\end{abstract}

\maketitle

\makeatletter
\newcommand{\manuallabel}[2]{\def\@currentlabel{#2}\label{#1}}
\makeatother
\manuallabel{fig:enlarge}{6}
\manuallabel{fig:mutualinformation}{7}
\manuallabel{fig:lineardatacollapse}{2}
\manuallabel{fig:logdatacollapse}{3}

The simple paradigmatic model of a particle hopping on a lattice in the presence of disorder significantly advanced our understanding of condensed matter systems.
It lead to the insight that a static disordered potential can lead to a complete absence of diffusion and hence conductance in an isolated quantum system.
This is known as \emph{Anderson localization} following its inception by Anderson \cite{Anderson:58} more than half a century ago \cite{Abrahams:10}.
The original formulation focused primarily on non-interacting systems and in the years following Anderson's work a complete picture was formed:
It is now known that non-interacting systems in one and two dimensions are localized for arbitrary disorder \cite{Lee:85,Stolz:2010}.
Anderson also conjectured that a closed system of interacting particles with sufficiently strong disorder would likewise localize and fail to equilibrate.
This conjecture was only recently put on a firmer theoretical footing in a seminal work by Basko, Aleiner and Altshuler \cite{Basko:06}.
This has led to a surge in interest in this phenomenon now known as \emph{many-body localization} (MBL).

The concept of MBL has been confirmed by a number of studies \cite{Oganesyan:07,Pal:10,Canovi:11,DeLuca:13,Kjall:14,Nandkishore:14,Luitz:15,Lev:2015,Rigol}, demonstrating that interacting systems can display a novel dynamical phase transition between a so called \emph{ergodic} and a \emph{many-body localized phase}.
The MBL phase is characterized by robust states protected by the extensively many (approximately) local integrals of motion which emerge \cite{Huse13a,Imbrie,serbyn13,chandran14,ros15}.
Many features of this MBL phase have since been explored.
For instance, it has been shown that in the MBL phase energy eigenstates typically have low entanglement entropy with respect to any bipartition, i.e., satisfy what is called an \emph{area law} \cite{Bauer:13,Friesdorf:14b,Area}.
This is in stark contrast to generic ergodic phases in which the entanglement entropy of eigenstates in the bulk of the spectrum exhibits an extensive volume law scaling.
For an initial pure product states, it has also been observed that in many-body localized systems, bipartite entanglement between two sectors of the system grows only logarithmically in time \cite{Znidaric:08,Bardarson:12,Vosk:13,Vosk:14,Abanin:13,Nanduri:14,Kjall:14,Friesdorf:14} until an extensive value is reached.
This differs notably from the usual power-law growth in ergodic systems, but also with the non-interacting case, in which a saturation to a constant is observed.
At the same time, many features of MBL are still unexplored and their broader connections unknown. 

In this work we go significantly beyond the previous approaches by applying a powerful and sensitive correlation measure to pin down and study the MBL transition. Our focus is on the time-averaged, dephased states that emerge from product initial states once the hopping part of the Hamiltonian is abruptly switched on. While fingerprints of the MBL transition are expected in the correlations of this dephased state, their utility depends strongly on the type of correlations considered. While the behavior of bipartite entanglement is a commonly used tool for characterization of phases by the condensed matter community, we go beyond this by employing a multi-partite correlation measure for mixed states. Quantum information theory classifies correlations in quantum states as \emph{classical correlations}, \emph{entanglement}, \emph{quantum correlations} and \emph{total correlations}, all of which have distinct physical interpretations and expose subtly different properties \cite{Modi:10, Modi:12}. Since we expect the inherently multipartite nature of correlations to play a role in the MBL transition, we argue that the \emph{total correlations} of the dephased state is both a meaningful and insightful quantity to investigate it.

Based on a precise condition for ergodicity we show that the total correlations in the dephased state exhibits a different scaling with the system size in ergodic and non-ergodic phases. In particular, in the disordered Heisenberg spin $1/2$ chain studied we find that the total correlations grow only logarithmically with the system size in the ergodic phase, while in the MBL phase the growth is linear. Studying via exact diagonalization the total correlations averaged over disorder realizations and pure product initial states, we show that in the crossover region between these two behaviors it develops a pronounced peak with a power-law decay with the disorder strength on either side, a key signature to identify and characterize the transition.
As a side remark, no peak is visible if instead of the total correlations we use the mutual information, a measure of bipartite correlations, between the left and right half of the system (see also Fig.~\ref{fig:mutualinformation} in the Supplemental Material), demonstrating that multi-partite correlations play an important role.

Additionally our study connects the problem of MBL with recent research on equilibration in coherently evolving quantum systems.
In the past decade this topic has seen an unprecedented revival of interest mainly due to spectacular experimental advances in cold atomic physics \cite{Polkovnikov:11,Eisert:14}.
In this platform, the coherent dynamics can be followed over long time scales. In fact, there is evidence for the first experimental realization of an MBL phase using cold atoms on optical lattices \cite{Schreiber:15}, adding further relevance to the work here. 

\paragraph{Total correlations.}
The MBL transition has been investigated with a variety of tools, from transport coefficients to level statistics.
A first diagnostic tool to capture real space correlations in quantum states is the growth of the entanglement entropy in the evolution of a product initial state \cite{Vosk:14,Bardarson:12}. Here, in view of the multipartite nature of correlations in interacting many-body systems, we sharpen this approach by employing the \emph{total correlations} $\T$  \cite{Modi:10,Modi:12,Vedral:02}. In order to define $\T$, we first introduce the relative entropy between two states $\rho$ and $\sigma$ defined by $S(\rho\|\sigma) \coloneqq -\tr(\rho\,\log_2 \sigma)-S(\rho)$, where $S(\rho) \coloneqq - \tr(\rho\,\log_2\rho)$ is the von Neumann entropy.
It is the quantum analogue of the Kullback-Leibler divergence and a very stringent measure of the distinguishability of two quantum states \cite{Audenaert:14} via a result known as quantum \emph{Stein's lemma}.
While not itself a metric, it still upper bounds the trace distance via \emph{Pinsker's inequality} $S(\rho\|\sigma) \geq \|\rho - \sigma \|_1^2/2$ \cite{Audenaert:14}, which captures the optimal distinguishability of quantum states with a single measurement.

We now introduce the total correlations $\T$:
Let $\mathcal{P}$ be the set of all product states of a $N$-partite quantum system, i.e., for spin systems, states of the form $\pi=\pi_{1}\otimes\pi_{2}\dots\otimes\pi_{N}$ (and the corresponding analogues for fermionic and bosonic systems).
The total correlations are then defined as the minimum relative entropy between the state and any product state, i.e.,
\begin{equation}
\T(\rho) \coloneqq \min_{\pi\in\mathcal{P}} S(\rho\|\pi) .
\end{equation}
It turns out that the unique product state which minimizes the relative entropy in the above definition is the product of the reduced states $\rho_m$ obtained from $\rho$ by tracing out all sites but the m-th , i.e., $\pi=\otimes^{N}_{m=1} \rho_{m}$ \cite{Modi:10}. This allows us to compute the quantity straightforwardly by making partial traces over the partitions of interest. The expression for the total correlations becomes
\begin{equation} \label{eq:totcor}
\T(\rho) = \sum_{m=1}^N S(\rho_{m}) - S(\rho) .
\end{equation} 
It is useful to point out that for $N=2$ the total correlations is equal to the \emph{mutual information}, which has the operational interpretation as the work required to erase the correlations in $\rho$ \cite{Groisman:05}.
If $\rho$ is a pure bipartite state, then the mutual information is equal to twice the entanglement entropy of $\rho$, i.e, $\T(\rho) = S(\rho_{1}) + S(\rho_{2})$.
We note that although the total correlations defined by 
Eq.~\eqref{eq:totcor} contains a contribution from the diagonal entropy studied in \cite{DeLuca:13} it also contains contributions from all marginal entropies and unlike the diagonal entropy is explitly related to the geometric picture of correlations in the state under investigation.

\paragraph{Quantum ergodicity, the diagonal ensemble and many-body localization.}
Leaving aside the problem of a proper definition of MBL, we take the complementary approach and start by defining a property that is a condition for rightfully calling a system \emph{ergodic}.
The \emph{ergodic hypothesis} in classical statistical physics states that ergodic systems explore their phase space uniformly such that the infinite time average and the microcanonical average should agree (making this precise is a subtle issue \cite{UffinkFinal}).
In quantum mechanics the time and the microcanonical average can agree exactly only for states that are evenly weighted coherent superpositions of all eigenstates in a microcanonical subspace \cite{vonNeumann:29}.
Hence, we require less and, informally speaking, take the standpoint that to call a system, i.e., a pair of Hamiltonian and initial state, \emph{ergodic} (as oppose to \emph{many body localized}) it should explore at least a constant fraction of the available Hilbert space.

Let us now turn this intuition into a clear cut definition.
The first step is to quantify the explored Hilbert space, we will do this based on the \emph{dephased} or \emph{time-averaged} state $\omega$.
For a fixed initial state $\rho$ and non-degenerate Hamiltonian $H$ we define
\begin{equation} \label{eq:timeav}
  \omega \coloneqq  \sum_n \ket {E_{n}}\bra {E_{n}} \,\rho\, \ket {E_{n}}\bra {E_{n}} = \lim_{\tau\to\infty} \frac{1}{\tau} \int^{\tau}_{0}dt\, \e^{-itH}\, \rho\, \e^{itH} ,
\end{equation}
where $\ket {E_{n}}$ are the eigenvectors of $H$.
This is often referred to as the \emph{diagonal ensemble}, as the off-diagonal elements are washed away by the time-average.
The dephased or time-averaged state is the unique state that maximizes the von Neumann entropy given all constants of motion \cite{Gogolin:11}.
If the expectation value of an observable equilibrates on average during the time evolution of a system, then the equilibrium expectation value can be computed from it \cite{vonNeumann:29,Eisert:14}.
What is more, under mild additional conditions on the Hamiltonian the following is true:
If the inverse purity $1/\tr(\omega^2)$ of the time averaged state, also called \emph{effective dimension} and \emph{participation ratio}, is high, expectation values of all sufficiently local observables equilibrate on average during the time evolution even if they were initially out of equilibrium \cite{Reimann:08,Linden:09,Polkovnikov:11,Eisert:14}.

The effective dimension, being a measure for the number of energy eigenstates that significantly contribute to the initial state \cite{Reimann:08,Linden:09}, can be interpreted as a measure for the explored Hilbert space fraction (as can other moments of the energy level occupation distribution, like $\tr(\omega^{2\,q})$ for $q \in \Z^+$ \cite{MultiFractal}).
Instead of demanding a large effective dimension for ergodicity we only demand the weaker property that $S(\omega) \geq \log(1/\tr(\omega^2))$ is large enough to call a system ergodic.

To identify a reasonable notion of being large enough we take inspiration from the theory of random states (although it is important to stress that we will not actually base any of the later calculations or numerics on Haar random states).
For a fixed Hamiltonian $H$ and randomly chosen unitarily invariant initial states $\rho(0)$ from the Haar measure on a microcanonical subspace of dimension $d$ one can show \cite[Eq.~(B6)]{Hayden:04} (compare also Refs.~\cite{lubkin,lloyd,facchi08,depasquale12}) that for some $C>0$
\begin{equation}
  \Pr\big(S(\omega) \leq \log_2(d/2)\big) \leq 4\, \exp(-C\, d/\log_2(d)^2).
\end{equation}
That is, random states typically explore at least half of the available Hilbert space in the sense that typically $S(\omega) \geq \log_2(d/2)$.

For our condition for ergodicity we relax this fraction of $1/2$ to a constant fraction of the available Hilbert space.
To make this meaningful we have to speak about families of systems of increasing system size $N$, specify what we mean by available Hilbert space and describe the class of initial states.
As is common in localization studies we take the subspace of dimension $d$ corresponding to a fixed filling $\eta\in[0,1]$ or magnetization $2\,\eta-1$ as the available Hilbert space.
We then consider initial states that are pure product states with definite local particle number or magnetization from that subspace, which can be thought of as ground states of appropriate ``easy'' Hamiltonians.
We say that a family of such systems should be considered ergodic only if most such product initial states explore at least a constant fraction of the fixed filling/magnetization subspace in the sense that for some $\lambda>0$ it holds that $S(\omega) \geq \log_2(\lambda\,d)$.
Note that this is less restrictive than demanding that $1/\tr(\omega) \geq \lambda\,d$, as $S(\omega) \geq -\log\tr(\omega)$.
For families of disordered systems we demand that the same condition is fulfilled with high probability also with respect to the disorder average.

\paragraph{Scaling of the total correlations.}
We now turn to demonstrating that the total correlations in the dephased state can be used to pin down and better understand the transition point from an ergodic to the MBL phase.
The key signature we exploit is the scaling of $T(\omega)$ with the system size $N$.
Inspecting Eq.~\eqref{eq:totcor} one might expect that the total correlations in the dephased state $\T(\omega)$ should generally scale extensively with $N$, i.e, for large $N$ one should have to leading order
\begin{equation} \label{eq:linearscaling}
  \T(\omega) \propto N ,
\end{equation}
as $\T(\omega)$ involves the sum $\sum_{m=1}^N S(\omega_{m})$ of the $N$ subsystem entropies.
Indeed, this is the behavior we find in the MBL phase of the model we consider below (see also Fig.~\ref{fig:lineardatacollapse} in the Supplemental Material).

If a family of disordered systems is ergodic however, then for some constant $\lambda > 0$, for most product initial states, and with high probability over the disorder average
\begin{equation} \label{eq:ergodicupperboundontotalcorrelations}
  \T(\omega) \leq \sum_{m=1}^N S(\omega_{m}) - \log_2(\lambda\,d) .
\end{equation}
For a quantum spin chain of local dimension $2$ at half filling $\eta=1/2$ the available Hilbert space dimension is $d = \binom{N}{N/2} = N! / \left(\tfrac{N}{2}!\right)^{2} \geq \sqrt{8\,\pi}\, \e^{-2}\, 2^N/\sqrt{N}$ and $S(\omega_{m}) \leq \log_2 2 = 1$, so that one finds at most the logarithmic scaling
\begin{equation} \label{eq:logscaling}
  \T(\omega) \leq \log_2(N)/2 - \log_2(\lambda\,\sqrt{8\,\pi}\,\e^{-2}) .
\end{equation}
This is what we observe in the ergodic phase of the model we consider.
One furthermore retains a logarithmic scaling for ergodic spin $1/2$ systems for all other constant fillings $\eta\in[0,1]$ if another mild condition is satisfied that can also be motivated from ergodicity, namely that for some $\lambda'>0$
\begin{equation} \label{eq:localaverageentropycondition}
  \sum_{m=1}^N S(\omega_m) \leq N\,s(\eta) + \lambda' \, \log_2 N
\end{equation}
where $s(x):= -x\,\log_2(x) - (1-x)\,\log_2(1-x)$ is the binary entropy function.
Eq.~\eqref{eq:localaverageentropycondition} says that the sum of the local entropies of the time averaged state should not grow much faster than one would expect for the given filling fraction $\eta$.
The generalized Stirling formula implies that
\begin{equation}
  \log_2 d = \log_2 \binom{N}{\eta\,N} \geq N\,s(\eta) - c(\eta) ,
\end{equation}
with $c(\eta) = 2 \log_2(\e) - \log_2(\eta) - \log_2(1-\eta)$.
Inserting this and Eq.~\eqref{eq:localaverageentropycondition} into \eqref{eq:ergodicupperboundontotalcorrelations} yields a logarithmic scaling with the system size:
\begin{equation}
  \T(\omega) \leq \lambda' \, \log_2 N - \log_2(\lambda) + c(\eta) 
\end{equation}

This sub-extensive scaling can also be understood intuitively:
The transport present in a ergodic systems correlates the different parts of the system to the extent that they appear, for most times during the evolution, so mixed that the distinguishability from the closest product state only grows logarithmically.

\paragraph{Model used for numerics.}
A model which is known to exhibit a crossover between an ergodic and a MBL phase is the Heisenberg spin chain with random field in the $z$ direction \cite{Pal:10}.
The Hamiltonian of this model is given by
\begin{equation} \label{eq:heisenbergchain}
  H =\sum^{N}_{i=1}\Big[J\,(\sigma^{i}_{x}\sigma^{i+1}_{x}+\sigma^{i}_{y}\sigma^{i+1}_{y})+J_z\,\sigma^{i}_{z}\sigma^{i+1}_{z}+h_{i}\sigma^{i}_{z}\Big]
\end{equation}
where the $h_{i}$ represent identically distributed static fields on each site $i$ uniformly distributed in the interval $[-h,h]$.
In what follows we adopt periodic boundary conditions and set $J_{z}=1$, so that a family of systems is completely characterized by the $XX$ type coupling constant $J$ and the disorder strength $h$.
For all values of the parameters, the model conserves the total magnetization $S_z$ along the $z$ direction, so in the numerics we have chosen the subspace with $S_z=0$, also referred to as half filling, i.e, $\eta=1/2$.
We take as our initial states all product eigenstates of the on-site part of the Hamiltonian $\sum_{i=1}^N \sigma^{i}_{z}$ from this subspace.
We then compute for each initial state the diagonal ensemble $\omega$ and $\T(\omega)$.
Averaging over all such initial states and disorder realizations yields $\Taverage(\omega)$.
The numerics were performed using standard libraries for matrix diagonalization.
We use $10,000$ disorder realizations for each disorder amplitude $|h|$ and system size $N$, except for the case of $N=16$ where $1000$ realizations per point were computed.

\begin{figure}[t]
\includegraphics[width=0.9\columnwidth]{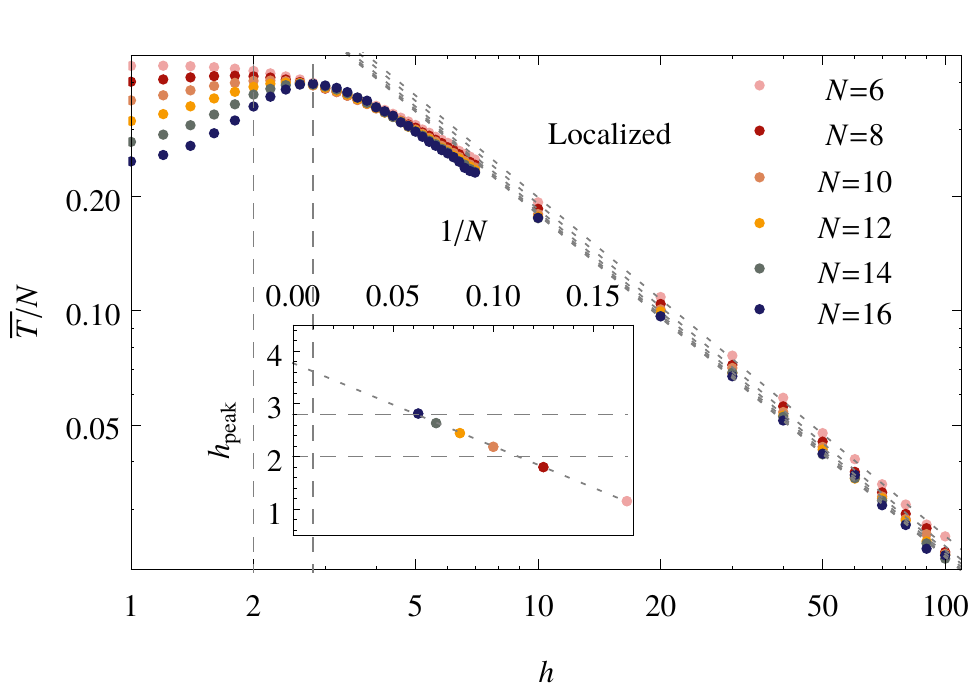}
\caption{\label{fig:powerlawdecay3} For high $h$, $\Taverage$ decays as a power-law. 
The dashed lines are fits to the data points with $h\geq10$ yielding exponents of $-0.9(1)$ (consistent with the expectation of $h^{-1}$ corrections).
The inset shows the position of the peaks from Fig.~\ref{fig:powerlawincrease} (see also Fig.~\ref{fig:enlarge} in the Supplemental Material).
The extrapolated position of the peak indicates the onset of MBL at around $h=3.8$. The fact that the different $\Taverage/N$ curves overlap for large enough $h$ indicates that $\Taverage$ scales essentially linearly with $N$ in this regime.}
\end{figure}

\begin{figure}[t]
\includegraphics[width=0.9\columnwidth]{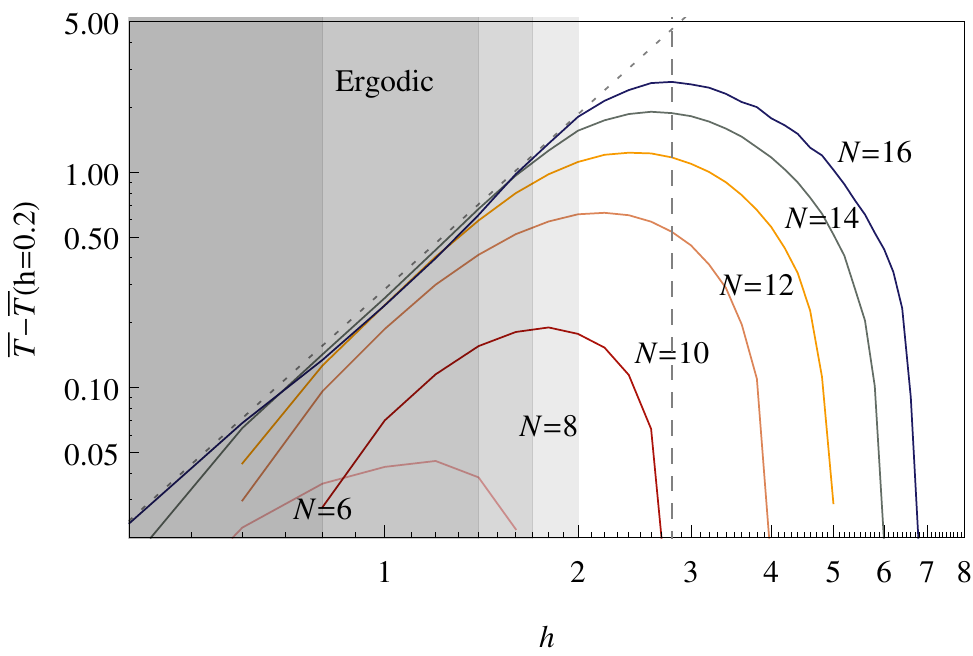}
\caption{Subtracting the values of $\Taverage$ for $h=0.2$ (which is just outside the integrable region around $h=0$) from $\Taverage$ one can see that for increasing system sizes the individual curves fall on top of each other in an increasingly large region of $h$ values (shaded regions) during the approach to the peak.
  This increase is well captured by a power-law with exponent $2.7(2)$ (dotted line guide to the eye $\propto h^{2.7}$).
  \label{fig:powerlawincrease}}
\end{figure}

\begin{figure}[t]
\includegraphics[width=0.9\columnwidth]{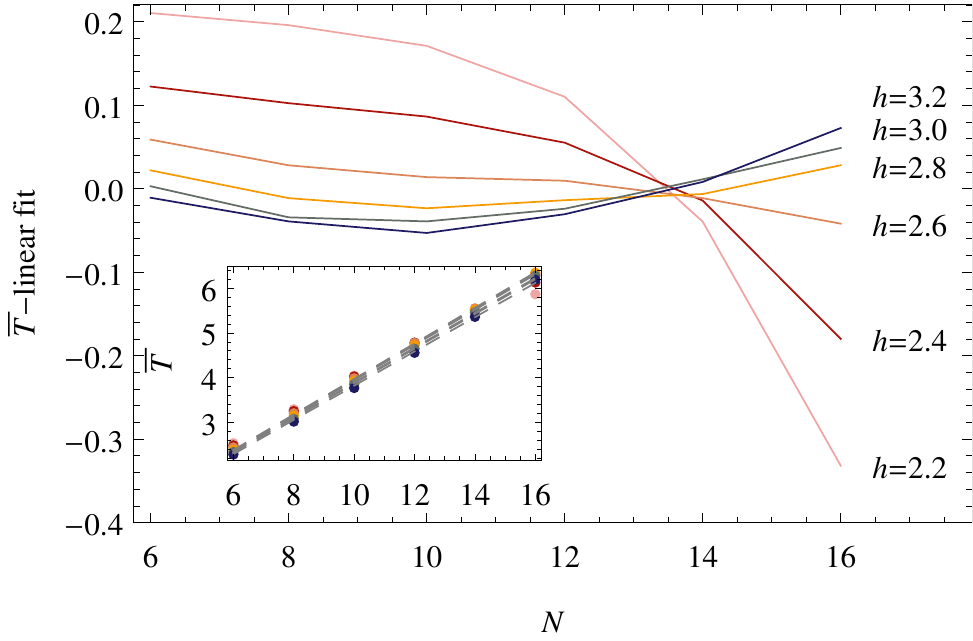}
\caption{\label{fig:breakdownoflinearscaling} 
  The difference of the average total correlations $\Taverage$ and the best possible linear fit for different values of $h$. The crossover from a nearly linear scaling to a sub-linear scaling clearly happens between $h=2.8$ and $h=2.4$. This result is robust against omitting data points for large or small values of $N$ and equally holds for affine fits instead of linear ones. This gives us high confidence in this result, which indicates the existence of an extended non-ergodic but not yet many-body localized region.The inset shows the data before subtracting and the fits.}
\end{figure}

\paragraph{Results, discussions and conclusions.}
We concentrate on the case $J=J_z=1$. For that case the MBL transition in the model \eqref{eq:heisenbergchain} was predicted to be $h_{c}\in [2,4]$ by Huse and Pal \cite{Pal:10}, with the best estimate based on energy resolved calculations being $h_c=3.72(6)$ \cite{Luitz:15} from an analysis of spectral statistics. Turning to the results we obtained, in all calculations performed we observe that the total correlations when plotted versus $h$ show an initial growth at low $h$ towards a maximum and then decrease monotonically at higher disorder with a power-law decay with an exponent of roughly $-0.9(2)$, i.e., $\Taverage \propto N\,h^{-0.9(2)}$ (see Fig.~\ref{fig:powerlawdecay3}).
This is consistent with the expectation of $h^{-1}$ corrections from perturbation theory and the behavior in the non-interacting case $J_z=0$ (see the Appendix).
The position of the maximum is size dependent, and can be extrapolated to be $h_c \approx 3.8$ in the thermodynamic limit. This is in excellent agreement with the best known approximation of the transition available in the literature \cite{Luitz:15}. In turn, by rescaling $T/N$ it appears that all curves collapse onto a single master curve for $h>h_c$ (see Fig.~\ref{fig:lineardatacollapse} in the Appendix).
Since on qualitative grounds we expect the many-body localized phase to be characterized by (i) linear scaling $T \propto N$, (ii) shrinking localization length as $h$ increases, it is natural to identify the MBL transition with the peak (see the inset of Fig.~\ref{fig:powerlawdecay3}).
The analysis for $h<h_c$ is more complex: scaling and data collapse for $T/\log(N)$, expected in an ergodic phase, are observed only for $h < 2.6(2)$ (see Fig.~\ref{fig:powerlawincrease} and \ref{fig:breakdownoflinearscaling}). For low disorder (up until $h \approx 2.0$ for the system sizes we can access), we see a power-law increase of $T$ with $h$ (see Fig.~\ref{fig:powerlawincrease}) with an exponent of about $2.7(2)$. While for $h<2$ the system is definitely in an ergodic phase (compare Fig.~\ref{fig:logdatacollapse} in the Supplemental Information), the analysis resented in Fig.~\ref{fig:breakdownoflinearscaling} suggest that ergodicity is is broken only around $h=2.6(2)$. This is consistent with an intermediate \emph{extended} yet \emph{non-ergodic} phase \cite{MultiFractal, Kamenev} before full MBL sets in. Due to the small system size finite size effects cannot be completely ruled out. Nevertheless, recent work has demonstrated an intermediate level statics \cite{intermediate}, implying non-ergodic extended states in precisely the crossover region indicated by our numerics. We believe that our work constitutes evidence of an intermediate non-ergodic region before the onset of MBL.

\paragraph{Conclusions.}
The numerical simulations performed together with our analytical arguments show that the total correlations in the diagonal ensemble signal both ergodicity breaking and the MBL transition in a quite spectacular way.
In standard critical systems it is known that the multi-partite correlations of the system rearrange as the system is pushed across an equilibrium phase transition \cite{Amico:08}.
Undoubtedly the transition from an ergodic to a MBL phase is a highly non-equilibrium phenomenon which is poorly understood at present. Our approach exposes how this transition goes along with a reorganization of correlations in the dephased state via significant change in scaling with $N$.
We expect this behavior to be generic and believe that the methodology outlined here is very promising to study MBL and ergodicity breaking phenomena in a variety of many-body quantum systems.
In a follow up study we will investigate the possible multi fractal nature of ergodicity breaking in a way inspired by Ref.~\cite{Kamenev, MultiFractal}. 

\paragraph{Acknowledgements.}
This work was partially supported by the COST Action MP1209. 
JE acknowledges support by the EU (SIQS, RAQUEL) and the ERC (TAQ),
CG acknowledges support by MPQ-ICFO, the Spanish Ministry Project FOQUS (FIS2013-46768-P), and the Generalitat de Catalunya (SGR 875). SRC acknowledges support from the ERC under the EU's Seventh Framework Programme (FP7/2007--2013)/ERC Grant Agreement no.~319286 Q-MAC. The authors 
acknowledge useful correspondence from M.~Rigol.

\section{Appendix}

\subsection{Total correlations in Anderson localization}
In this section, we complement the results presented in the main text by discussing the total correlations for the time averaged state in the random $XX$ model, with Hamiltonian 
\begin{equation}
  H =\sum^{N}_{i=1}[J\sigma^{i}_{x}\sigma^{i+1}_{x}+J\sigma^{i}_{y}\sigma^{i+1}_{y}+h_{i}\sigma^{i}_{z}].
\end{equation}
The above model is equivalent to the model in the text for $J_z=0$, and serves as a simple testbed in which very large system sizes can easily be probed.
Specifically, for suitable boundary conditions, the above model is equivalent to the free fermionic model 
\begin{equation}
  H = f^\dagger M f,
\end{equation}
which we take as the basis for our analysis of \emph{non-interacting disordered models}. Here,
$f=(f_1,\dots, f_N)^T$ is the collection of \emph{free fermionic annihilation operators} 
of $N$ fermionic modes. In the above quadratic form, the kernel $M$ is given by
\begin{equation}
	M:= - 2\,J\, \circula(0,1,0,\dots, 0,1) - 2\, \diag(h_1,\dots, h_N),
\end{equation}
again with $(h_1,\dots, h_N)^T$ drawn uniformly random from $[-h,h]^N$. This real symmetric matrix can be diagonalized as 
\begin{equation}
	M=O D O^T,\,\, O\in O(N),
\end{equation}
with $D$ being real and diagonal.
More generally, unitary transformations $U\in U(N)$ from one set of fermionic operators to another one can be allowed for, 
and in all what follows, orthogonal transformations can be replaced by unitaries. For the present purposes,
this is unnecessary, however.

Gaussian states $\rho$ of systems of massive fermions (no Majorana fermions are considered) can be captured in terms of \emph{correlation matrices} $C(\rho)\geq 0$, with entries
\begin{equation}
	C(\rho)_{j,k}:= {\rm tr}(f_j^\dagger f_k\rho),\quad j,k \in \{1,\dots,N\}.
\end{equation}
For a given correlation matrix $C(\rho)$ of such a state $\rho$ the correlation matrix of the state $\sigma$ expressed in the basis in which the Hamiltonian is diagonal is given by 
\begin{equation}
	C(\sigma)= O^T\,C(\rho)\,O. 
\end{equation}
It is easy to see that 
the correlation matrix
of the infinite time average $\omega$ of the initial state $\rho$ represented by $C(\rho)$
is then 
\begin{equation}
	C(\omega) = O\,\Pi(O^T\,C(\rho)\,O)\,O^T, 
\end{equation}
where $\Pi$ is the map that projects a matrix onto its main diagonal.

The \emph{ground state correlation matrix} can again be expressed in terms of the kernel $M$ of the
Hamiltonian form, as long as $\{0\}\not\in \text{spec}(M)$: 
Then the ground state $\rho$ is unique and has the correlation matrix
\begin{equation}
  C(\rho) = O\,\sgn(\diag(M))\,O^T ,
\end{equation}
as again can be verified by expressing the Hamiltonian in the appropriate basis.

\begin{figure}[tb]
\includegraphics[width=0.95\columnwidth]{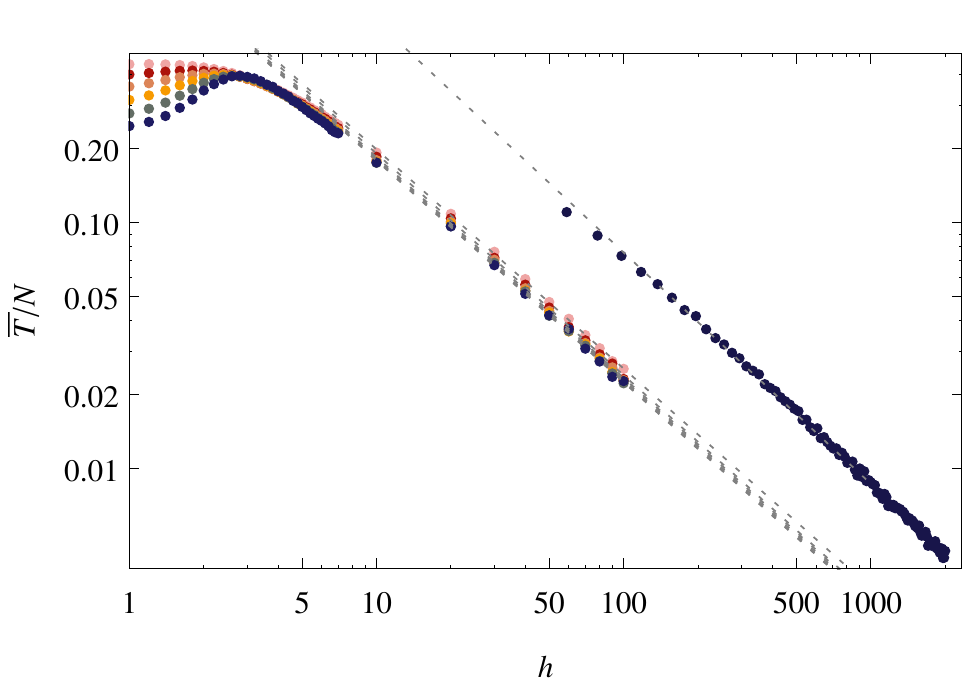}
\caption{\label{fig:freemodels} 
  Data points on the right are the averaged total correlations of the free model with $J_z=0$.
  The behavior is again well captured by a power law with exponent $-0.9(1)$.
  For comparison on the left is the data shown in Fig.~\ref{fig:powerlawdecay3} of the main text.
  Dotted lines are power law fits.
}
\end{figure}

Entropies of Gaussian states can be computed from their correlation matrices. 
Making again use of the binary entropy function one finds that 
any such Gaussian state $\rho$ of $N$ modes 
with correlation matrix $1\geq C(\rho)\geq 0$ has the von-Neumann entropy
\begin{equation}
	S(\rho) =\tr (s(C(\rho))),
\end{equation}	
as can be seen by exploiting suitable orthogonal mode transformations
and the unitary invariance of the von-Neumann entropy on the level of quantum states.
Hence, the total correlations of the time averaged state $\omega$
are found to be 
\begin{equation}
  \begin{split}
    T(\rho) &= \tr\big( s( \Pi(O\,\Pi(O^T\,C(\rho)\,O)\,O^T)\big)\\
    &-  \tr\big(s(\Pi(O^T\,C(\rho)\,O))\big) .
  \end{split}	  
\end{equation}
That is to say, both entropies of the infinite time averaged state and its reductions can be conveniently computed.
In this way, once the correlation matrix has been identified, the total correlations measure can be immediately obtained. Drawing i.i.d.\ random vectors $(h_1,\dots, h_N)^T$ uniformly from $[-h,h]^N$ as in the main text, one can
very clearly identify the power law decay of the total correlations.
Numerically, system sizes of $N=100$, can easily be accommodated in this way, finding that $\log \Taverage/N \,\log(h)$ can be
well fitted with an affine function, reflecting a power law, again with exponent $-0.9(1)$ (see Fig.~\ref{fig:freemodels}).

\subsection{Further plots}
In the following we present some additional plots for the Heisenberg spin chain with random field in Eq.~\eqref{eq:heisenbergchain} of the main text with 
with $J=J_Z=1$.
\vfill
\begin{figure}[h]
\includegraphics[width=0.9\columnwidth]{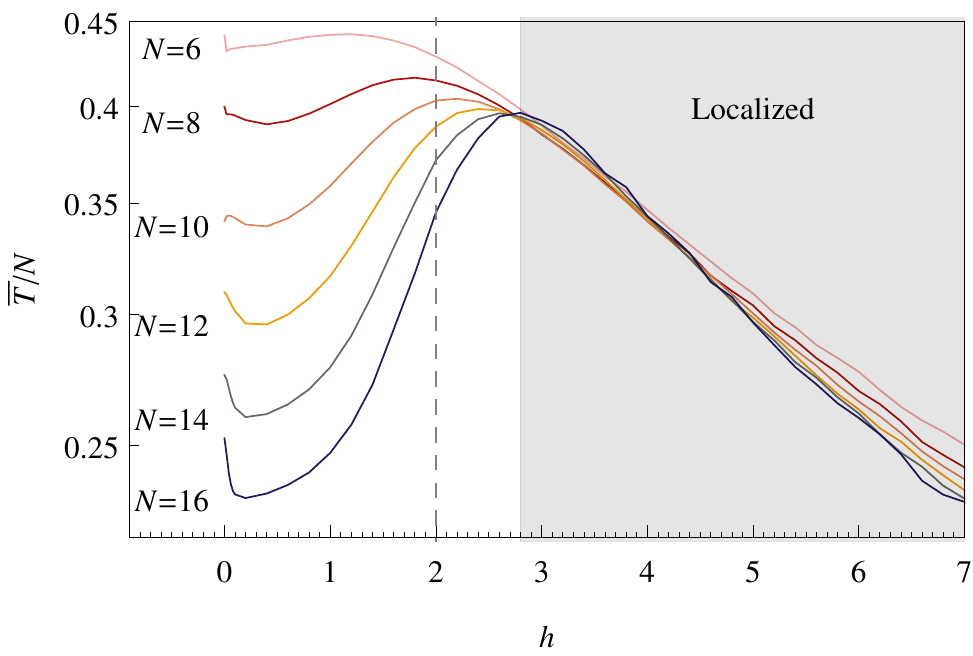}
\caption{\label{fig:lineardatacollapse}
Semi logarithmic plot of the intensive averaged total correlations for the Hamiltonian in Eq.~\eqref{eq:heisenbergchain} with $J=J_z=1$ versus $h$ for system sizes between $6$ and $16$.
The merging of the different $\Taverage/N$ curves for large $h$ demonstrates that we can reliably determine that $\Taverage$ exhibits the linear scaling predicted in Eq.~\eqref{eq:linearscaling} with the system size $N$ down to $h \approx 2.6$.}
\end{figure}
\vfill
\begin{figure}[h]
\includegraphics[width=0.95\columnwidth]{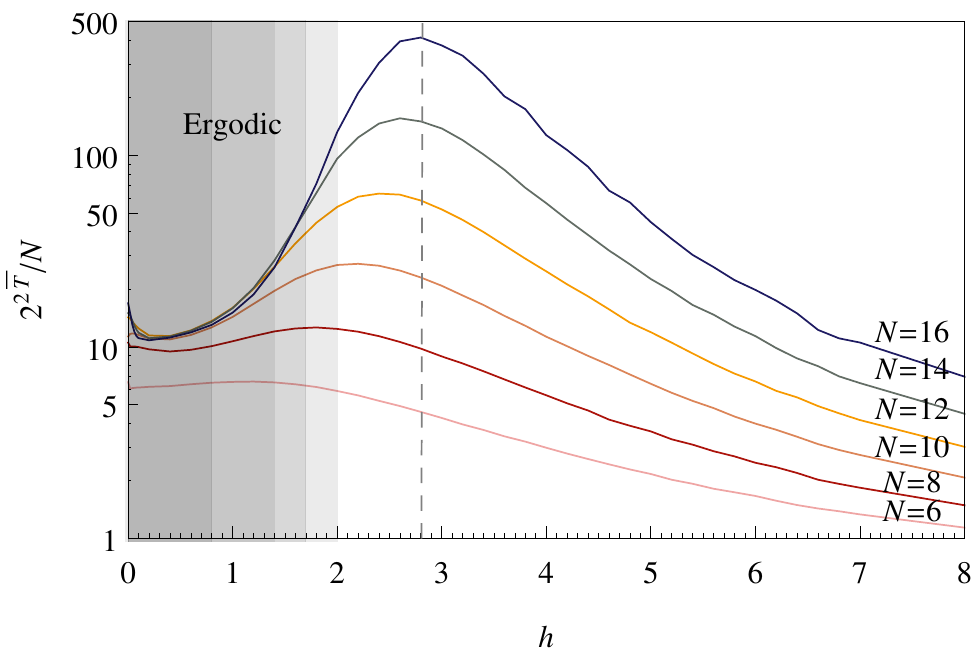}
\caption{\label{fig:logdatacollapse} Semi logarithmic plot of $2^{2 \overline{T}}/N$ versus $h$ for system sizes between $6$ and $16$.
The merging of the different curves demonstrates that we can reliably determine that $\Taverage$ exhibits the logarithmic scaling predicted in Eq.~\eqref{eq:logscaling} in the main text with the system size $N$ up to at least $h \approx 2.0$. The more elaborate analysis performed in Fig.~\ref{fig:breakdownoflinearscaling} of the main text even pushes this to $h=2.6(2)$.}
\end{figure}

\begin{figure}[h]
\includegraphics[width=0.95\columnwidth]{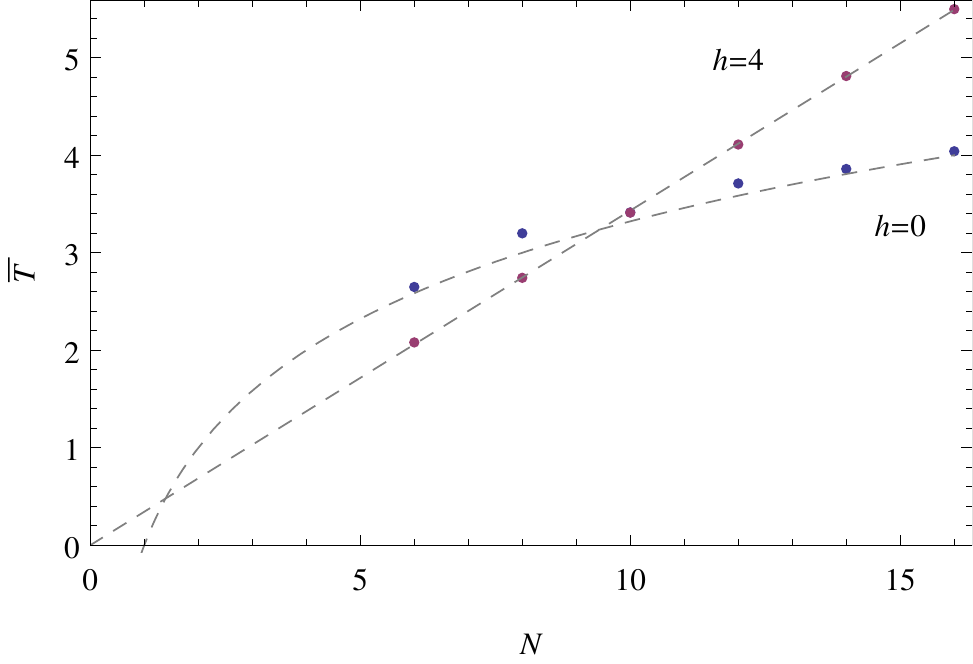}
\caption{\label{fig:integrable} 
Scaling of the averaged total correlations $\Taverage$ for the translation invariant case without disorder $h=0$ and in the MBL phase at $h=4$.
  The dashed lines are linear fit to the $h=4$ data and the graph of $\log_2 N$ (no fit) for comparison.
  As in the ergodic phase we find a logarithmic scaling even at $h=0$.
}
\end{figure}

\begin{figure}[h]
\includegraphics[width=0.95\columnwidth]{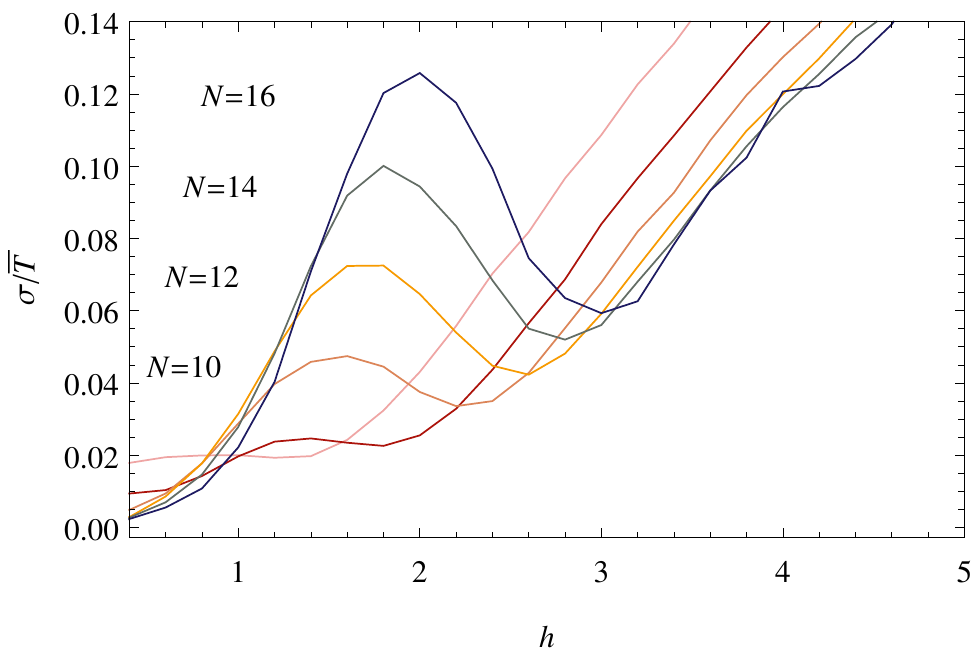}
\caption{\label{fig:variance} 
 The standard deviation $\sigma$ of $\T$ with respect to both the disorder average and the average over product initial states divided by $\Taverage$ as a function of $h$.
  Well within a phases either (MBL or ergodic) $\T$ is self averaging. However, if closer to the transition the quantity is affected by rare events.
  We expect the region around the peaks, in which $\sigma/\Taverage$ scales linearly with $N$, shifts further to the right, into the region of the phase transition with increasing system size.}
\end{figure}

\begin{figure}[h]
\includegraphics[width=0.95\columnwidth]{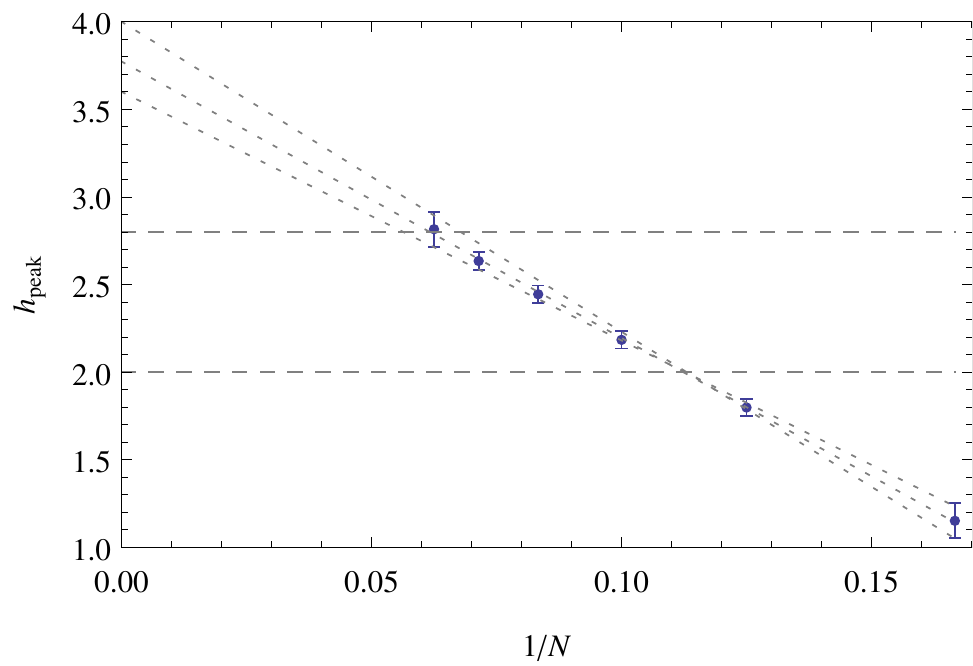}
\caption{\label{fig:enlarge} 
 The figure shows the position of the peaks in Fig.~\ref{fig:powerlawincrease} of the main text and is an enlarged version of the inset of Fig.~\ref{fig:powerlawdecay3} of the main text.}
\end{figure}

\begin{figure}[h]
\includegraphics[width=0.95\columnwidth]{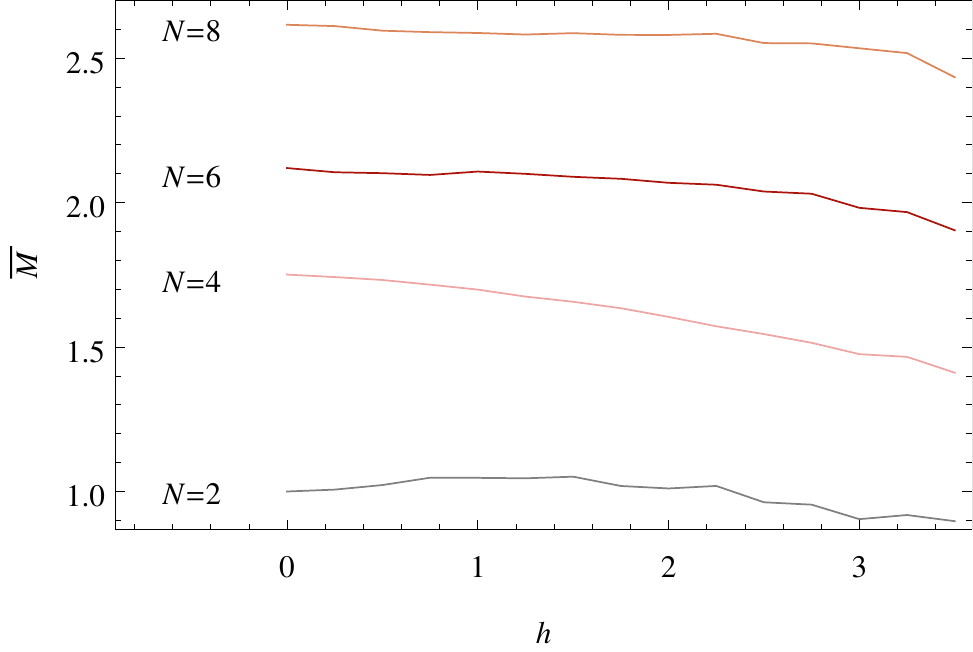}
\caption{\label{fig:mutualinformation} 
Compared to the total correlations, the mutual information $M(\rho) \coloneqq S(\rho_{[1,N/2]}) + S(\rho_{[N/2+1,N]}) - S(\rho)$, when computed in the time averaged state and averaged in the same way, appears to be mostly featureless in the parameter range of the localization transition.
This supports our point that the multipartite nature of the total correlations is the reason for why it signals the transition so well.
The plot shows an average over 100 disorder realizations.}
\end{figure}

\cleardoublepage 

\end{document}